# Efficient Nash Computation in Large Population Games with Bounded Influence

**Michael Kearns**
Department of Computer and Information Science
University of Pennsylvania
Philadelphia, Pennsylvania
mkearns@cis.upenn.edu

**Yishay Mansour**
School of Computer Science
Tel Aviv University
Tel Aviv, Israel
mansour@cs.tau.ac.il

## Abstract

We introduce a general representation of large-population games in which each player's influence on the others is centralized and limited, but may otherwise be arbitrary. This representation significantly generalizes the class known as congestion games in a natural way. Our main results are provably correct and efficient algorithms for computing and learning approximate Nash equilibria in this general framework.

## 1 INTRODUCTION

We introduce a compact representation for single-stage matrix games with many players. In this representation, each player is influenced by the actions of all others, but only through a global *summarization function*. Each player's payoff is an arbitrary function of their own action and the value of the summarization function, which is determined by the population joint play. This representation of large-population games may be exponentially more succinct that the naive matrix form, and here we prove that vast computational savings can be realized as well. A natural example of such games is voting scenarios (a special case in which the summarization function is both linear and symmetric), where each player's payoff depends only on their own vote and the outcome of the popular election, but not on the details of exactly how every citizen voted. Certain large-population financial market and auction models are also natural candidates. As discussed below, summarization games generalize a number of existing representations in the game theory literature, such as congestion games.

We make the natural assumption of *bounded influence* — that is, that no player can unilaterally induce arbitrarily large changes in the value of the summarization function. (Voting is a simple example of bounded influence.) Under only this assumption and a bound on the derivatives of the private individual payoff functions (both of which appear to be necessary), we give an algorithm for efficiently computing approximate Nash equilibria, which interestingly always outputs *pure* approximate equilibria. We also prove that a simple variant of distributed smoothed best-response dynamics will quickly converge to (learn) an approximate equilibrium for any *linear* summarization function. These algorithms run in time polynomial in the number of players and the approximation quality parameter, and are among the few examples of provably efficient Nash computation and learning algorithms for broad classes of large-population games.

## 2 RELATED WORK

A closely related body of work is the literature on games known as *congestion games* (Rosenthal [1973]) or *exact potential games* (Monderer and Shapley [1996]), which are known to be equivalent. In congestion games and their generalizations, players compete for a central resource or resources, and each player's payoff is a (decreasing) function of the number of players selecting the resources. An example is the well-studied *Santa Fe Bar problem*, where patrons of a local bar receive positive payoff if their numbers are sufficiently low, negative payoff if they exceed capacity, and players who stay home receive 0 payoff. Single-resource congestion games can be viewed as summarization games in which the global summarization is *symmetric* — that is, dependent only on the total number of players selecting the resource. In the current work we allow the summarization function to be both non-symmetric and non-linear, but our results can also be viewed as a contribution to the congestion game literature. While a fair amount is understood about the mathematical properties of equilibria in congestion games (such as the existence of pure equilibria), and there has been a great deal of recent experimental simulation (see, for example, Greenwald et al. [2001]), there seems to be relatively little work providing provably efficient and correct algorithms for computing and learning equilibria.

We also view the proposed representation and algorithms as complementary to recent work on compact undirected graphical models for multi-player games (Kearns et al.

[2001], Littman et al. [2002], Vickrey and Koller [2002]). While those works emphasize large-population games in which each player is strongly influenced by a small number of others, the current work focuses on games in which each player is weakly influenced by all others. This is analogous to the two main cases of tractable inference in Bayesian networks, where the polytree algorithm provides an algorithm for sparse networks, and variational algorithms yield approximate inference in dense networks with small-influence parametric CPTs.

## 3 DEFINITIONS AND NOTATION

We begin with the standard definitions for multiplayer matrix games. An $n$-player, two-action [1] game is defined by a set of $n$ matrices $M_i$ ($1 \leq i \leq n$), each with $n$ indices. The entry $M_i(x_1, \ldots, x_n) = M_i(\vec{x})$ specifies the payoff to player $i$ when the joint action of the $n$ players is $\vec{x} \in \{0,1\}^n$. Thus, each $M_i$ has $2^n$ entries. We shall assume all payoffs are bounded in absolute value by 1.

The actions 0 and 1 are the *pure strategies* of each player, while a *mixed* strategy for player $i$ is given by the probability $p_i \in [0, 1]$ that the player will play 0. For any joint mixed strategy, given by a product distribution $\vec{p}$, we define the expected payoff to player $i$ as $M_i(\vec{p}) = \mathbf{E}_{\vec{x} \sim \vec{p}}[M_i(\vec{x})]$, where $\vec{x} \sim \vec{p}$ indicates that each $x_j$ is 0 with probability $p_j$ and 1 with probability $1 - p_j$.

We use $\vec{p}[i : p'_i]$ to denote the vector which is the same as $\vec{p}$ except in the $i$th component, where the value has been changed to $p'_i$. A *Nash equilibrium* for the game is a mixed strategy $\vec{p}$ such that for any player $i$, and for any value $p'_i \in [0,1]$, $M_i(\vec{p}) \geq M_i(\vec{p}[i : p'_i])$. (We say that $p_i$ is a *best response* to the rest of $\vec{p}$.) In other words, no player can improve their expected payoff by deviating unilaterally from a Nash equilibrium. The classic theorem of Nash [1951] states that for any game, there exists a Nash equilibrium in the space of joint mixed strategies.

We will also use a straightforward definition for approximate Nash equilibria. An $\epsilon$-*Nash equilibrium* is a mixed strategy $\vec{p}$ such that for any player $i$, and for any value $p'_i \in [0,1]$, $M_i(\vec{p}) + \epsilon \geq M_i(\vec{p}[i : p'_i])$. (We say that $p_i$ is an $\epsilon$-*best response* to the rest of $\vec{p}$.) Thus, no player can improve their expected payoff by more than $\epsilon$ by deviating unilaterally from an approximate Nash equilibrium.

As in Kearns et al. [2001], our goal is to introduce a natural new representation for large multiplayer games that is considerably more compact than the classical tabular form, which grows exponentially in the number of players. However, rather than succinctly capturing games in which each player has a small number of possibly strong "local" influences, our interest here is at the other extreme — where each player is influenced by *all* of the others in a large pop-ulation, but no single player has a dominant influence on any other.

The first element of this new representation is a population *summarization function* that we shall denote $\mathcal{S}(\vec{x})$. As discussed in the introduction, a natural example might be the *voting* summarization function $\mathcal{S}(\vec{x}) = (1/n) \sum_{i=1}^{n} x_i$, but here we explicitly allow $\mathcal{S}$ to be asymmetric and non-linear. We shall assume without loss of generality that the range of $\mathcal{S}$ is $[0, 1]$. The central idea is that in a multiplayer game in which the joint pure play is $\vec{x}$, the payoffs to any player $i$ will be a function of only his own action $x_i$, and the two values $\mathcal{S}(\vec{x}[i : 0])$ and $\mathcal{S}(\vec{x}[i : 1])$. We view these two values as *summarizing* for each player all that they need to know about the joint behavior in order to decide how to act. Note that one of these two values (namely, $\mathcal{S}(\vec{x}[i : x_i])$) is simply $\mathcal{S}(\vec{x})$.

A natural question is why we provide each player with the two values $\mathcal{S}(\vec{x}[i : 0])$ and $\mathcal{S}(\vec{x}[i : 1])$, rather than simply the single value $\mathcal{S}(\vec{x})$. The reason is that in many natural cases, the single-value model may lead to situations in which players do not have sufficient global information to determine the effects of their own actions, or even compute their best responses, since the value $\mathcal{S}(\vec{x})$ alone may not determine the effect on $\mathcal{S}$ of changing $x_i$. As an example, suppose that $\mathcal{S}(\vec{x})$ reports the fraction of players that are playing the majority value, without specifying whether the majority value is 0 or 1, and that $i$ is a player whose payoff increases with the value of $\mathcal{S}$ (that is, $i$ is a consensus-builder). Then for any given value of $\mathcal{S}(\vec{x})$, $i$ cannot determine whether he should change his play from $x_i$ in order to build greater consensus. If he is provided with both $\mathcal{S}(\vec{x}[i : 0])$ and $\mathcal{S}(\vec{x}[i : 1])$, he can directly see the impact on $\mathcal{S}$ of his own actions (but not that of others), and can at least always compute his own best response to any fixed pure joint play $\vec{x}$. We note, however, that the results we will give shall render this distinction between receiving one or both values largely a technicality, and that the reader may informally think of the players as receiving the single summarizing value $\mathcal{S}(\vec{x})$.

Before moving on to the payoff functions for the individual players, we first discuss restrictions on $\mathcal{S}$ that we shall assume. Recall that our goal is to model settings in which every player's payoff may depend on every other player's move, but in a way that the influences are bounded. For a fixed summarization function $\mathcal{S}$, we formally define the *influence* of player $i$ by

$$\tau_i = \max_{\vec{x} \in \{0,1\}^n} \{|\mathcal{S}(\vec{x}[i : 0]) - \mathcal{S}(\vec{x}[i : 1])|\}.$$

The influence measures the greatest change player $i$ can ever (unilaterally) effect in the summarization function. We say that the influence of $\mathcal{S}$ is bounded by $\tau$ if $\tau_i \leq \tau$ for all players $i$. In keeping with our motivation, we shall be studying the computational benefits that can arise from multiplayer games with bounded influence summarization functions. Note that since we assume the range of $S$ is $[0, 1]$, the influence is always bounded by 1; however, if there are $n$ players, *in many natural bounded-influence settings the maximum influence will be on the order of $1/n$ (as in voting), or at least some function diminishing with $n$.*

---

[1] We describe our results for the two-action case. The generalization to multiple actions is straightforward.

Finally, we discuss the individual payoff functions. Here we simply assume that each player $i$ possesses separate payoff functions for their two actions, $\mathcal{F}_0^i$ and $\mathcal{F}_1^i$. If the pure actions of the other $n-1$ players are given by the vector $\vec{x}$ ($x_i$ is irrelevant here) the payoff to $i$ for playing 0 is defined to be $\mathcal{F}_0^i(\mathcal{S}(\vec{x}[i:0]))$, and the payoff to player $i$ for playing 1 is defined to be $\mathcal{F}_1^i(\mathcal{S}(\vec{x}[i:1]))$. Thus, for any joint play $\vec{x}$, each player is told what values their two actions will yield for the population summarization function, and has private payoff functions indicating their own reward for each resulting value. We assume that the $\mathcal{F}_b^i$ are real-valued functions mapping $[0, 1]$ to $[0, 1]$. We shall also assume that all the payoff functions are continuous and have bounded derivatives.

Note that even though the summarization function $S$ has bounded influence, and thus a player's action can have only bounded effect on the payoffs to others, it can have dramatic effect on his *own* payoffs, since $\mathcal{F}_0^i$ and $\mathcal{F}_1^i$ may assume quite different values for any mixed strategy (despite the bounds on their derivatives). We feel this is a natural model for many large-population settings, where subjective (private) regret over actions may be unrelated to the (small) influence an individual has over global quantities. For instance, a staunch democrat might personally find voting for a republican candidate quite distasteful, even though this individual action might have little or no effect on the overall election. It is their private payoff functions that makes individuals "care" about their actions in a large population where the global effects of any single player are negligible.

We shall assume throughout that the summarization function $S$ and all the payoff functions $\mathcal{F}_b^i$ can be efficiently computed on any given input; formally, we will assume such a computation takes unit time. Thus, the tuple $\mathcal{G} = (S, \{(\mathcal{F}_0^i, \mathcal{F}_1^i)\}_{i=1}^n)$, which we shall call a *(large population) summarization game*, is a representation of an $n$-player game that may be considerably more compact than its generic matrix representation. We say that $\mathcal{G}$ is a $(\tau, \rho)$-summarization game if the influence of $S$ is bounded by $\tau$, and the derivatives of all $\mathcal{F}_b^i$ are bounded by $\rho$.

Two final remarks on the definition of a summarization game are in order. First, note that the representation is entirely general: by making the summarization and payoff functions sufficiently complex, any $n$-player game can be represented. If $S$ outputs enough information to reconstruct its input (for example, by computing a weighted sum of its inputs, where the weight of bit $x_i$ is $2^{-i}$), and the payoff functions simply interpolate the values of the original game matrices for player $i$, the original game is exactly represented. However, in such cases we will not have small influence and derivatives, and our results will naturally degrade. It is only for bounded influence and derivative games, which seem to have wide applicability, that our results are interesting. Second, we note that if we view the summarization function as being defined for every input length $n$ (as in voting) and fixed, and the continuous payoff functions as being fixed, then summarization games naturally represent games with an arbitrarily large or growing number of players, and our results will shed light on computing and learning equilibria in the limit of large populations.

The results in this paper describe efficient algorithms for computing and learning approximate Nash equilibria in summarization games, and provide rates of convergence as a function of summarization influence, payoff derivatives, and population size. We now turn to the technical development.

## 4 COMPUTING EQUILIBRIA

The first of our two main results is an efficient algorithm for computing approximate Nash equilibria in bounded-influence summarization games:

**Theorem 1** *There is an algorithm **SummNash** that takes as input any $\epsilon > 0$ and any $(\tau, \rho)$-summarization game $\mathcal{G} = (S, \{(\mathcal{F}_0^i, \mathcal{F}_1^i)\}_{i=1}^n)$ over $n$ players, and outputs an $O(\epsilon + \tau\rho)$-Nash equilibrium for $\mathcal{G}$. Furthermore, this approximate equilibrium will be a pure (deterministic) joint strategy. The running time of **SummNash** is polynomial in $n$, $1/\epsilon$, and $\rho$.*

Before presenting the proof, let us briefly interpret the result. First, the accuracy parameter $\epsilon$ is an input to the algorithm, and thus can be made arbitrarily small at the expense of the polynomial dependence on $1/\epsilon$ of the running time. As for the $\tau\rho$ term in the approximation quality, it is natural to think of the derivative bound $\rho$ as being a fixed constant, while the influence bound $\tau$ is some diminishing function of the number of players $n$ — that is, individuals have relatively smooth payoffs independent of population size, while their individual influence on the summarization function shrinks as the population grows. Under such circumstances, Theorem 1 yields an algorithm that will compute arbitrarily good approximations to equilibria as the population increases.

The proof of Theorem 1 and the associated algorithm will be developed in a series of lemmas. Our first step is to approximate the continuous, bounded-derivative individual payoff functions $\mathcal{F}_b^i$ by piecewise constant (step) functions $\hat{\mathcal{F}}_b^i$. For a given resolution $\alpha$ (to be determined by the analysis), we divide $[0, 1]$ into the $\alpha$-*intervals* $[0, \alpha), [\alpha, 2\alpha), [2\alpha, 3\alpha), \ldots$. Denote the $k$th such interval as $I_k = [k\alpha, (k+1)\alpha)$. We define the approximation $\hat{\mathcal{F}}_b^i$ to be constant over any $\alpha$-interval $I$. Specifically, for any $z \in I_k$, $\hat{\mathcal{F}}_b^i(z) = \mathcal{F}_b^i(k\alpha)$. Since the derivative of the $\mathcal{F}_b^i$ is bounded by $\rho$, we have $|\mathcal{F}_b^i(z) - \hat{\mathcal{F}}_b^i(z)| \leq \rho\alpha$ for all players $i, b \in \{0, 1\}$, and $z \in [0, 1]$. In the sequel, we shall refer to $\hat{\mathcal{G}} = (\mathcal{S}, \{(\hat{\mathcal{F}}_0^i, \hat{\mathcal{F}}_1^i)\}_{i=1}^n)$ as the $\alpha$-*approximate summarization game* for $\mathcal{G}$.

We first show that the bounded derivatives of the payoff functions translates to a Lipschitz-like condition on the approximate payoff functions.

**Lemma 2** *For all $z, z' \in [0, 1]$,*

$$|\hat{\mathcal{F}}_b^i(z) - \hat{\mathcal{F}}_b^i(z')| \leq (\rho|z - z'| + 2\rho\alpha)\theta(z - z')$$

*where we define $\theta(z - z') = 0$ if $z = z'$ and $\theta(z - z') = 1$ otherwise.*

**Proof:** Clearly the difference is 0 if $z = z'$. If $z \neq z'$ we have

$$|\hat{\mathcal{F}}_b^i(z) - \hat{\mathcal{F}}_b^i(z')| \leq |\mathcal{F}_b^i(z) - \mathcal{F}_b^i(z')| + 2\rho\alpha \leq \rho|z - z'| + 2\rho\alpha$$

where the first inequality comes from the approximation quality, and the second from the bound on the derivatives of the $\mathcal{F}_b^i$. □

The following straightforward lemma translates the quality of approximate Nash equilibria in $\hat{\mathcal{G}}$ back to the original game $\mathcal{G}$.

**Lemma 3** *Let $\vec{p}$ be any $\gamma$-Nash equilibrium for the $\alpha$-approximate summarization game $\hat{\mathcal{G}}$. Then $\vec{p}$ is a $(2\rho\alpha + \gamma)$-Nash equilibrium for $\mathcal{G}$.*

**Proof:** Since $\vec{p}$ is a $\gamma$-Nash equilibrium for $\hat{\mathcal{G}}$, each player $i$ is playing a $\gamma$-best response. The rewards in $\mathcal{G}$ can change by at most $\rho\alpha$ for each action, which implies that the change to a new best response is at most $2\rho\alpha + \gamma$. □

We next give a lemma that will simplify our arguments by letting us define (approximate) best responses solely in terms of the single global summarization value $\mathcal{S}(\vec{x})$, rather the multiple local values $\mathcal{S}(\vec{x}[i:b])$ for each $i$ and $b$. We start with the following definition.

**Definition 1** *Let $\hat{\mathcal{G}}$ be the $\alpha$-approximate summarization game, and let $\vec{p}$ be any mixed strategy. We define for player $i$ the* single-value apparent best response *in $\hat{\mathcal{G}}$ as*

$$\hat{a}_i(\vec{p}) = \mathrm{argmax}_{b \in \{0,1\}}\{\mathbf{E}_{\vec{x} \sim \vec{p}}[\hat{\mathcal{F}}_b^i(\mathcal{S}(\vec{x}))]\}.$$

Thus, $\hat{a}_i(\vec{p})$ is the apparent best response for $i$ in $\hat{\mathcal{G}}$ to $\vec{p}$ if $i$ ignores the effect of his own actions on the summarization function. We now show that this apparent best response in $\hat{\mathcal{G}}$ is in fact an approximate best response in $\hat{\mathcal{G}}$.

**Lemma 4** *Let $\hat{\mathcal{G}}$ be the $\alpha$-approximate summarization game. Let $\vec{p}$ be any mixed strategy. Then $\hat{a}_i(\vec{p})$ is a $(\tau\rho + 2\rho\alpha)$-best response for player $i$ to $\vec{p}$ in $\hat{\mathcal{G}}$.*

**Proof:** For any pure strategy $\vec{x}$ and any $b \in \{0, 1\}$, we have

$$|\mathcal{S}(\vec{x}) - \mathcal{S}(\vec{x}[i:b])| \leq \tau|b - x_i|$$

due to the bound on influence. Note that the right-hand side of this inequality is $\tau$ if $b \neq x_i$ and 0 if $b = x_i$. By Lemma 2, we have

$$|\hat{\mathcal{F}}_b^i(\mathcal{S}(\vec{x})) - \hat{\mathcal{F}}_b^i(\mathcal{S}(\vec{x}[i:b]))| \leq (\tau\rho + 2\rho\alpha)|b - x_i|.$$

Taking expectations under $\vec{p}$ gives

$$\left|\mathbf{E}_{\vec{x} \sim \vec{p}}[\hat{\mathcal{F}}_b^i(\mathcal{S}(\vec{x}))] - \mathbf{E}_{\vec{x} \sim \vec{p}}[\hat{\mathcal{F}}_b^i(\mathcal{S}(\vec{x}[i:b]))]\right| \leq (\tau\rho + 2\rho\alpha)|b - x_i|.$$

Now if

$$\hat{a}_i(\vec{p}) \neq \mathrm{argmax}_{b \in \{0,1\}}\{\mathbf{E}_{\vec{x} \sim \vec{p}}[\hat{\mathcal{F}}_b^i(\mathcal{S}(\vec{x}[i:b]))]\}$$

(that is, $\hat{a}_i(\vec{p})$ is not already a *true* best response to $\vec{p}$ for $i$ in $\hat{\mathcal{G}}$), then the inequality above implies it is a $(\tau\rho + 2\rho\alpha)$-best response. □

Now note that by construction, if $I_k$ is any $\alpha$-interval, and $\vec{x}$ and $\vec{x}'$ are any two pure strategies such that $\mathcal{S}(\vec{x}) \in I_k$ and $\mathcal{S}(\vec{x}') \in I_k$ (that is, both vectors give a value of the summarization function lying in the same $\alpha$-interval), then $\hat{a}_i(\vec{x}) = \hat{a}_i(\vec{x}')$, because the approximate payoff functions $\hat{\mathcal{F}}_b^i$ do not change over $I_k$. Furthermore, the action $\hat{a}_i(\vec{x})$ is an approximate best response for $i$ in $\hat{\mathcal{G}}$ by Lemma 4. In other words, in $\hat{\mathcal{G}}$, we have reduced to a setting in which the (approximate) best response of all players can be viewed solely as a function of the $\alpha$-interval $I_k$ in which $\mathcal{S}(\vec{x})$ lies, and not on the details of $\vec{x}$ itself.

For any $\alpha$-interval $I_k$, let us define

$$\vec{\mathrm{BR}}(I_k) = \langle \hat{a}_1(\vec{x}), \ldots, \hat{a}_n(\vec{x}) \rangle$$

where $\vec{x}$ is any vector such that $\mathcal{S}(\vec{x}) \in I_k$. Thus, $\vec{\mathrm{BR}}(I_k)$ is the vector of (apparent) best responses of the players in $\hat{\mathcal{G}}$ when the value of the summarization function falls in $I_k$. This best response itself gives a value to the summarization function, which we define as $\mathcal{V}(I_k) = \mathcal{S}(\vec{\mathrm{BR}}(I_k))$. We can extend this definition to view $\mathcal{V}$ as a mapping from $[0, 1]$ to $[0, 1]$ (rather than from $\alpha$-intervals to $[0, 1]$) by defining $\mathcal{V}(x)$ to be $\mathcal{V}(I_k)$, where $x \in I_k$. In Figure 1 we provide a sample plot of a hypothetical $\mathcal{V}$ that we shall refer to for expository purposes.

The purpose of the definition of $\mathcal{V}$ is made clear by the next lemmas. The intuition is that those places where $\mathcal{V}$ "crosses" the diagonal line $y = x$ are indicative of approximate Nash equilibria. We begin with the easier case in which $\mathcal{V}$ crosses the diagonal during one of its constant-valued horizontal segments, marked as the point $A$ in Figure 1.

**Lemma 5** *Let $I_k$ be an $\alpha$-interval such that $\mathcal{V}(I_k) \in I_k$. Then $\vec{\mathrm{BR}}(I_k)$ is a $(\tau\rho + 2\rho\alpha)$-Nash equilibrium for $\hat{\mathcal{G}}$.*

**Proof:** Let $\vec{x} = \vec{\text{BR}}(I_k)$. Since $\mathcal{V}(I_k) = \mathcal{S}(\vec{x}) \in I_k$, every player $i$ is playing $x_i = \hat{a}_i(\vec{x})$, and thus by Lemma 4, a $(\tau\rho + 2\rho\alpha)$-best response to $\vec{x}$. □

We next examine the case where Lemma 5 does not apply. First we establish a property of the function $\mathcal{V}$.

**Lemma 6** *If for every $\alpha$-interval $I_k$, $\mathcal{V}(I_k) \notin I_k$, then there exists a $k$ such that $\mathcal{V}(I_{k-1}) > k\alpha > \mathcal{V}(I_k)$.*

**Proof:** For $k=0$ we have $\mathcal{V}(I_0) > 0$, and for $\ell = 1/\alpha - 1$ (the last interval) we have $\mathcal{V}(I_\ell) < 1$. Therefore there has to be a $k$ for which the lemma holds. □

In other words, if Lemma 5 does not apply, there must be two consecutive intervals whose $\mathcal{V}$-values "drop" across the diagonal. This case is illustrated in Figure 1 by the vertical dashed line containing the point $B$.

**Lemma 7** *Let $k$ be such that $\mathcal{V}(I_{k-1}) > k\alpha > \mathcal{V}(I_k)$. Then there is a pure strategy $\vec{x}$ which is a $(3\tau\rho + 6\rho\alpha)$-Nash equilibrium in $\hat{\mathcal{G}}$.*

**Proof:** Let $\vec{y} = \vec{\text{BR}}(I_{k-1})$ and $\vec{z} = \vec{\text{BR}}(I_k)$. Let $t$ be the number of indices $i$ for which $y_i \neq z_i$. Define the sequence of $t$ vectors $\vec{x}^1, \ldots, \vec{x}^t$ such that $\vec{x}^1 = \vec{y}$, $\vec{x}^t = \vec{z}$, and for every $j = 1, \ldots, t$, $\vec{x}^{j+1}$ is obtained from $\vec{x}^j$ by flipping the next bit $i$ such $y_i \neq z_i$. Thus, in each $\vec{x}^j$, bits that have the same value in $\vec{y}$ and $\vec{z}$ are unaltered, while bits that differ in $\vec{y}$ and $\vec{z}$ flip exactly once during the "walk" from $\vec{x}$ to $\vec{z}$.

The intuition is that if we can find any vector on the walk which gives a value to $\mathcal{S}$ falling in or near the interval $I_k$, it must be an approximate Nash equilibrium, since players whose best response in this neighborhood of $\mathcal{S}$-values may be strongly determined (that is, those for which $y_i = z_i$) are properly set throughout the walk, while the others may be set to either value (since they are nearly indifferent in this neighborhood, as evidenced by their switching approximate best response values from $I_{k-1}$ to $I_k$). In Figure 1, the points along the vertical line that include $B$ conceptually show the different values of $\mathcal{V}$ during the walk.

Now for each $j$ we define $v_j = \mathcal{S}(\vec{x}^j)$. Due to the bounded influence of $\mathcal{S}$, we have that $|v_j - v_{j+1}| \leq \tau$ for all $j = 1, \ldots, t$. This implies that for some value $\ell$, we have $|v_\ell - k\alpha| \leq \tau$. (In Figure 1, point $B$ corresponds to the vector $\vec{x}^\ell$ on the walk whose $\mathcal{V}$-value comes closest to $k\alpha$, and thus constitutes the desired approximate Nash equilibrium.)

We now show that $\vec{x}^\ell$ is an approximate Nash equilibrium. Consider player $i$, and assume that $x_i^\ell = 0$, but that the best response for $i$ is actually 1 (the other cases are similar). If $v_\ell \in I_k$, then $x_i^\ell$ is a $(\tau\rho + 2\alpha\rho)$-best response by Lemma 4. Otherwise, $v_\ell \in I_{k-1}$. Since $0 = x_i^\ell$ is the apparent best response for $k\alpha$, we have

$$\hat{\mathcal{F}}_1^i(k\alpha) \leq \hat{\mathcal{F}}_0^i(k\alpha).$$

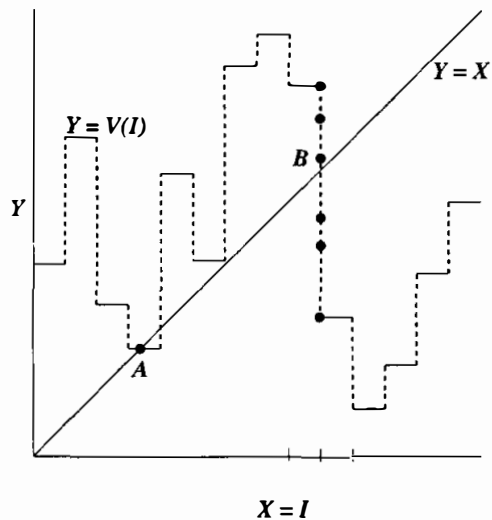

$X = I$

Figure 1: Hypothetical plot of the function $\mathcal{V}$ over the $\alpha$-intervals $I_k$. We view the $x$-axis as being both a continuum of individual points and a discrete sequence of $\alpha$-intervals. $\mathcal{V}$ generally begins above the diagonal line $y = x$, and ends up below the diagonal, and thus must cross the diagonal at some point. The point labeled $A$ is an example of a horizontal crossing as covered by Lemma 5. The column of points including the point labeled $B$ is an example of a vertical crossing as covered by Lemma 6. Each point in this column indicates a value of $\mathcal{S}$ realized on the vector "walk" discussed in the proof of Lemma 6, while the point $B$ itself is the value of $\mathcal{S}$ nearest the diagonal in this walk. The analysis establishes that (at least) one of these two crossing cases must occur.

We now bound the difference in the reward to player $i$ due to playing action 0 rather than 1 in response to $\vec{x}^\ell$, which is

$$\hat{\mathcal{F}}_1^i(\mathcal{S}(\vec{x}^\ell[x_i:1])) - \hat{\mathcal{F}}_0^i(\mathcal{S}(\vec{x}^\ell[x_i:0])).$$

This difference is bounded by

$$\hat{\mathcal{F}}_1^i(v_\ell) - \hat{\mathcal{F}}_0^i(v_\ell) + \tau\rho + 2\rho\alpha$$

since $v_\ell = \mathcal{S}(\vec{x}^\ell)$, and by Lemma 2. Again by Lemma 2, we have

$$|\hat{\mathcal{F}}_b^i(v_\ell) - \hat{\mathcal{F}}_b^i(k\alpha)| \leq \tau\rho + 2\rho\alpha.$$

Putting it all together gives

$$\hat{\mathcal{F}}_1^i(\mathcal{S}(\vec{x}^\ell[x_i:1])) - \hat{\mathcal{F}}_0^i(\mathcal{S}(\vec{x}^\ell[x_i:0])) \leq 3\tau\rho + 6\rho\alpha.$$

Therefore every player is playing a $(3\tau\rho + 6\rho\alpha)$-best response. □

Lemma 5 covers the case where $\mathcal{V}$ crosses the diagonal horizontally, and Lemma 7 the complementary case where it does not. Either case leads to the efficient computation of a $(3\tau\rho + 6\rho\alpha)$-Nash equilibrium for $\hat{\mathcal{G}}$, which by Lemma 3 is a $(3\tau\rho + 8\rho\alpha)$-Nash equilibrium for the original game $\mathcal{G}$. Setting $\epsilon = 8\rho\alpha$ or $\alpha = \epsilon/(8\rho)$ completes the proof of Theorem 1.

A description of the algorithm **SummNash** detailed in the above analysis is provided in Figure 2.

---

**Algorithm SummNash**
**Inputs:** An $n$-player $(\tau, \rho)$-summarization game $\mathcal{G} = (\mathcal{S}, \{(\mathcal{F}_0^i, \mathcal{F}_1^i)\}_{i=1}^n)$, and an approximation parameter $\epsilon$. We assume that $\mathcal{S}$ and each payoff function $\mathcal{F}_b^i$ are provided as black-box subroutines that cost unit computation time to call.
**Output:** A $(3\tau\rho + \epsilon)$-Nash equilibrium for $\mathcal{G}$.

1. $\alpha \leftarrow \epsilon/8\rho$.

2. For each player $i = 1, \ldots, n$, and $b \in \{0, 1\}$, construct the $\alpha$-approximate payoff functions $\hat{\mathcal{F}}_b^i$: for every $\alpha$-interval $I_k = [k\alpha, (k+1)\alpha)$, $\hat{\mathcal{F}}_b^i(z) = \mathcal{F}_b^i(k\alpha)$ for all $z \in I_k$. Note that this requires $1/\alpha$ evaluations of each $\mathcal{F}_b^i$.

3. Construct the mapping $\mathcal{V}$ by setting $\mathcal{V}(I_k) = \mathcal{S}(\vec{\text{BR}}(I_k))$ for every $\alpha$-interval $I_k$.

4. For each $\alpha$-interval $I_k$, $k = 1, \ldots, 1/\alpha$, check if $\mathcal{V}(I_k) \in I_k$. If so, then output $\vec{\text{BR}}(I_k)$.

5. For each $\alpha$-interval $I_k$, $k = 1, \ldots, 1/\alpha$, check if $\mathcal{V}(I_{k-1}) > k\alpha > \mathcal{V}(I_k)$. If so:

   (a) Let $A$ be the set of player indices whose play is the same in $\vec{\text{BR}}(I_{k-1})$ and $\vec{\text{BR}}(I_k)$.
   (b) $B \leftarrow \{1, \ldots, n\} - A$.
   (c) $t \leftarrow |B|$.
   (d) $\vec{x}^1 \leftarrow \vec{\text{BR}}(I_{k-1})$.
   (e) For $j = 1, \ldots, t$, let $\vec{x}^j$ be obtained from $\vec{x}^{j-1}$ by flipping the bit corresponding to the next index in $B$. Note that $\vec{x}^t = \vec{\text{BR}}(I_k)$.
   (f) For $j = 1, \ldots, t$, if $|\mathcal{S}(\vec{x}^j) - k\alpha| \leq \rho$, output $\vec{x}^j$.

---

Figure 2: Algorithm **SummNash**.

## 5 LEARNING IN LINEAR GAMES

In this section, we propose and analyze a distributed learning algorithm for the players of a game in which the summarization function $\mathcal{S}$ is linear, that is, $\mathcal{S}(\vec{x}) = \sum_{i=1}^n w_i x_i$. It is easily verified that in such a case that each influence $\tau_i = w_i$. Our main result is that the learning algorithm converges to an approximate Nash equilibrium for the summarization game $\mathcal{G}$. The analysis will rely heavily on the tools developed in Section 4.

The learning algorithm for each player is rather simple, and can be viewed as a variant of smoothed best response dynamics (Fudenberg and Levine [1999]). First of all, if $\mathcal{G} = (\mathcal{S}, \{(\mathcal{F}_0^i, \mathcal{F}_1^i)\}_{i=1}^n)$ is the original (linear) summarization game, each player $i$ will instead use the $\alpha$-approximate payoff functions $\hat{\mathcal{F}}_b^i$ described in Section 4. Note that these approximations can be computed privately by each player.

We shall use $\vec{p}^t$ to denote the joint mixed strategy of the population at time $t$ in the learning algorithm. In our learning model, at each round $t$, the expected value of the summarization function

$$\mu_t = \mathcal{S}(\vec{p}^t) \equiv \mathbf{E}_{\vec{x} \sim \vec{p}^t}[\mathcal{S}(\vec{x})]$$

is broadcast to all players [2].

---

[2] In the full paper we will examine the variant in which a pure strategy $\vec{x}$ is *sampled* according to $\vec{p}^t$, and all players receive only the value $\mathcal{S}(\vec{x})$. The analysis and results will be similar due to large-deviation bounds.

One natural learning algorithm would call for each player $i$ to update $p_i^t$ in the direction of their apparent best response $\hat{a}_i(\vec{p}^t)$. This apparent best response involves expectations over $\vec{p}^t$, and therefore requires each player to know this distribution to compute $\hat{a}_i(\vec{p}^t)$ exactly. However, let us extend the definition of apparent best response to apply to means rather than complete distributions:

$$\hat{a}_i(\mu_t) = \text{argmax}_{b \in \{0,1\}} [\hat{\mathcal{F}}_b^i(\mu_t)].$$

We can view $\hat{a}_i(\mu_t)$ to be the *approximate* apparent best response for $i$ to $\vec{p}^t$ under the simplifying assumption that $\vec{p}^t$ generates a distribution of $\mathcal{S}(\vec{x})$ that is sharply peaked near its mean value $\mu_t$. Before describing the learning algorithm, we state without proof a standard large deviation bound that we shall use in the analysis, establishing that this assumption is warranted in large populations.

**Lemma 8** *Let* $\hat{\mathcal{G}} = (\mathcal{S}, \{(\hat{\mathcal{F}}_0^i, \hat{\mathcal{F}}_1^i)\}_{i=1}^n)$ *be the $\alpha$-approximate summarization game derived from the $(\tau, \rho)$-summarization game $\mathcal{G} = (\mathcal{S}, \{(\mathcal{F}_0^i, \mathcal{F}_1^i)\}_{i=1}^n)$. Let $\vec{p}$ be any mixed strategy, and let $\mu = \mathcal{S}(\vec{p})$. Then for each player $i$, $\hat{a}_i(\mu)$ is a $\psi$-best response to $\vec{p}$ in $\hat{\mathcal{G}}$, for $\psi = O(\rho \sqrt{\sum_{i=1}^n \tau_i^2} \log(1/\sqrt{\sum_{i=1}^n \tau_i^2}))$.*

This lemma actually holds regardless of whether $\mathcal{S}$ is linear, but we shall only apply it to the linear case. Thus, for example, if all weights in $\mathcal{S}$ are bounded by some constant $C$ times $1/n$ ($C = 1$ in the case of straight voting), we have $\psi = O(\rho \log(n)/\sqrt{n})$. Thus, as before we have improved approximations with larger populations. Note, however, that $\sqrt{\sum_{i=1}^n \tau_i^2} \geq \tau = \max_i\{\tau_i\}$.

We now describe the learning algorithm, which has a learning rate parameter $0 < \beta < 1$ and a "stopping" parameter $\delta$. We view $\beta$ as a small fixed constant, and for the analysis will require that $\beta < \alpha$. At time $t$, each player $i$ updates their mixed strategy *slightly* in the direction of $\hat{a}_i(\mu_t)$:

$$p_i^{t+1} = (1-\beta)p_i^t + \beta\hat{a}_i(\mu_t).$$

If we define $\vec{BR}(\mu) \equiv \langle \hat{a}_1(\mu), \ldots, \hat{a}_n(\mu)\rangle$, the global vector update resulting from these distributed updates is

$$\vec{p}^{t+1} = (1-\beta)\vec{p}^t + \beta\vec{BR}(\mu_t).$$

If $|p_i^{t+1} - p_i^t| \leq \delta$ for *all* $i$, then the learning algorithm is terminated; otherwise updates continue for all players.

We will refer to this distributed learning algorithm as **SummLearn**($\delta$). Note that if $\delta > 0$, the algorithm requires a single additional bit of global information at each step, or the ability for players to "broadcast" if their update was greater than $\delta$. For **SummLearn**(0), no such mechanism is necessary, as all players continue updating forever. Below we shall consider both cases, because while **SummLearn**(0) is more decentralized and therefore more natural, we can make a slightly stronger statement in the $\delta > 0$ case. Note that **SummLearn**(0) is simply an approximate and smoothed version of best-response dynamics (Fudenberg and Levine [1999]) — each player simply moves their mixed strategy slightly in the direction of the best response to a sharply peaked distribution of $S$ with the broadcast mean.

For the analysis, we begin by writing:

$$\begin{aligned}
\mu_{t+1} &= S(\vec{p}^{t+1}) \\
&= S((1-\beta)\vec{p}^t + \beta\vec{BR}(\mu_t)) \\
&= (1-\beta)S(\vec{p}^t) + \beta S(\vec{BR}(\mu_t)) \\
&= (1-\beta)\mu_t + \beta S(\vec{BR}(\mu_t)) \\
&= (1-\beta)\mu_t + \beta \mathcal{V}(I_k)
\end{aligned}$$

where we define $I_k$ to be the $\alpha$-interval containing $\mu_t$, we have used the linearity of $S$, and $\mathcal{V}$ is as defined in Section 4. The above implies

$$\mu_{t+1} - \mu_t = \beta(\mathcal{V}(I_k) - \mu_t).$$

In other words, as long as $\mu_t < \mathcal{V}(I_k)$, the distributed updates cause the mean to increase, and as long as $\mu_t > \mathcal{V}(I_k)$, they cause the mean to decrease. Viewed graphically in Figure 3, the learning dynamics cause the mean to move towards the crossing points of the function $\mathcal{V}$ analyzed in Section 4. As before, there are two distinct cases of crossing. We now analyze the convergence to the crossings.

Since $\beta < \alpha$ by choice, we always have $|\mu_{t+1} - \mu_t| \leq \alpha$. This implies that we can only move to adjacent $\alpha$-intervals:

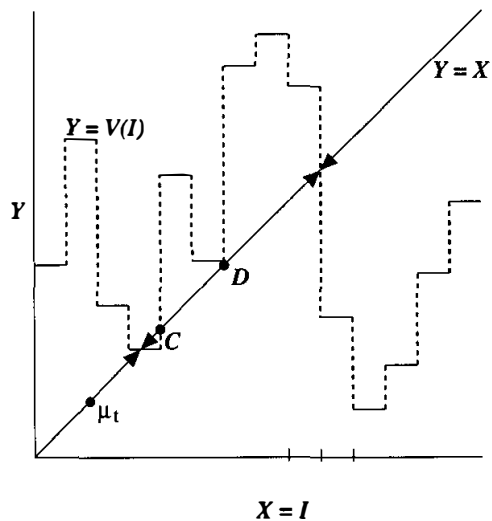

Figure 3: Replication of Figure 1, with arrows added along the diagonal to illustrate the dynamics of **SummLearn**($\delta$). We can think of the distributed learning process as generating the movement of the mean $\mu_t = S(\vec{p}^t)$ along the diagonal. As indicated by the arrows, at each point $\mu_t$, if $\mathcal{V}$ is above the diagonal, then the learning dynamics increase $\mu_t$, and if $\mathcal{V}$ is below the diagonal they decrease $\mu_t$. Convergence is to crossing points of $\mathcal{V}$. At the point labeled $C$ the dynamics reverse direction, and the point labeled $D$ indicates a neighborhood where the dynamics do not reverse, but may slow down considerably, since $\mathcal{V}$ is near the diagonal.

if $\mu_t \in I_k$, then $\mu_{t+1} \in I_{k-1} \cup I_k \cup I_{k+1}$. Let us say that learning has *converged* at time $T$ if there exists a vector $\vec{p}^*$ such that for all $t \geq T$ we have $\vec{p}^t = \vec{p}^*$. We first show that if learning converges, it must be to an approximate Nash equilibrium. This covers two possible outcomes of the learning dynamics: convergence to a horizontal crossing point, or convergence to a point on a horizontal segment that comes very close to the diagonal (see Figure 3).

**Lemma 9** *Suppose learning has converged at time $T$ to $\vec{p}^*$. Then $\vec{p}^*$ is an $O(\psi + \rho\delta + \rho\alpha)$-Nash equilibrium in $\mathcal{G}$.*

**Proof:** Let $\mu^* = S(\vec{p}^*)$. Since the learning has converged, we have $|\hat{a}^i(\mu^*) - p_i^*| \leq \delta$ for every $i$. By Lemmas 2 and 8, $p_i^*$ must be a $(\psi + \rho\delta + 2\rho\alpha)$-best response in $\hat{\mathcal{G}}$, and thus by Lemma 3, $\vec{p}^*$ is a $(\psi + \rho\delta + 4\rho\alpha)$-Nash equilibrium in $\mathcal{G}$. □

Now for any $\alpha$-interval $I_k$, we call a maximal sequence of consecutive time steps $t$ with $\mu_t \in I_k$ a *visit* to interval $I_k$. The *duration* of a visit is the number of steps in the visit.

**Lemma 10** *The duration of any visit to any $\alpha$-interval $I_k$ is bounded by $(1/\beta)\ln(1/\delta)$.*

**Proof:** As long as $\mu_t \in I_k$, $\vec{BR}(\mu_t)$ is unchanged, and therefore $\|\vec{p}^t - \vec{BR}(\mu_t)\|_\infty$ is reduced by a factor of $1-\beta$ at each step. Solving $(1-\beta)^m \leq \delta$ for $m$ yields the desired bound. □

The next claim follows immediately from Lemma 10 and the Pigeonhole Principle.

**Lemma 11** *Assume that the mean $\mu_t$ never visits any $\alpha$-interval twice. Then* **SummLearn**$(\delta)$ *converges after at most* $(1/\alpha)(1/\beta)\ln(1/\delta)$ *steps.*

The following lemma handles both the case that learning never converges, and the case that some interval is visited twice.

**Lemma 12** *Suppose that at some time $T$, $\mu_t$ makes a second visit to some $\alpha$-interval $I_k$. Then for all $t > T'$, the mixed strategies $\vec{p}^t$ are all $O(\psi + \rho\tau + \rho\delta + \rho\alpha)$-Nash equilibria for $\mathcal{G}$, where $T' < T + (1/\beta)\ln(1/\delta)$.*

**Proof:** Let $I_k$ be the first $\alpha$-interval visited twice by $\mu_t$, and that $T$ is the first step of the second visit. Since $\beta < \alpha$, at time $T - 1$ we had either $\mu_{T-1} \in I_{k-1}$ or $\mu_{T-1} \in I_{k+1}$; we assume the latter case without loss of generality. Note that since $I_k$ is the first revisited $\alpha$-interval, $\mu_t$ is monotonically increasing while $\mu_t \in I_k$, and monotonically decreasing while $\mu_t \in I_{k+1}$. Thus, for all $t > T$, we will have $\mu_t \in I_k \cup I_{k+1}$.

Consider a player $i$ such that $\hat{a}_i(I_k) = \hat{a}_i(I_{k+1})$. After at most $(1/\beta)\ln(1/\delta)$ time steps we will have $|\vec{p}_i^t - \hat{a}_i(I_k)| \leq \delta$. Therefore, as in the proof of Lemma 9, by time $T'$ player $i$ will play an $O(\psi + \rho\alpha + \rho\delta)$-best response in $\hat{\mathcal{G}}$ for all $t > T$. For the other players $j$ such that $\hat{a}_j(I_k) \neq \hat{a}_j(I_{k+1})$, any action is an $O(\psi + \rho\tau + \rho\alpha)$-best response for all $t > T$, since $\mu_t \in I_k \cup I_{k+1}$. □

For the case $\delta > 0$, we thus have the following theorem, which together with Theorem 15 below constitutes the second of our main results.

**Theorem 13** *After at most $O((1/\alpha)(1/\beta)\log(1/\delta))$ steps,* **SummLearn**$(\delta)$ *plays an $O(\psi + \rho\tau + \rho\delta + \rho\alpha)$-Nash equilibrium for all subsequent time steps.*

As in algorithm **SummNash**, we can make the term $\rho\delta + \rho\alpha$ smaller than any desired $\epsilon$ by choosing $\delta = \epsilon/2\rho$ and $\alpha = \epsilon/2\rho$, with the resulting polynomial dependence on $1/\epsilon$ in the running time. This leaves us with the uncontrollable term $\psi + \rho\tau$. Again, as we have discussed, in many reasonable large-population games we expect these influence terms to vanish as $n$ becomes large. Also, note that given that we require the learning rate $\beta < \alpha$, there is no benefit to setting $\beta$ much smaller than this, and thus the choice $\beta = \alpha/2$ yields an overall convergence time of $O((1/\alpha^2)\log(1/\delta))$.

We now analyze **SummLearn**(0). Here we cannot expect to upper bound the time it will take to converge to an approximate Nash equilibrium — technically, if the interval $I_k$ is such that $|\mathcal{V}(I_k) - \alpha k| \leq \delta$, $\mu_t$ might stay in $I_k$ for $\Theta(\log 1/\delta)$ steps (see Figure 3). Since $\delta$ can be arbitrary small, this time cannot be bounded. However, we can show that any time $\mu_t$ is "near the diagonal", the players are playing an approximate Nash equilibrium. This implies that we can bound the number of time steps in which they are *not* playing an approximate Nash equilibrium.

**Lemma 14** *Consider a visit of $\mu_t$ to interval $I_k$ in the time interval $t \in \{t_1, \ldots, t_2\}$. For all times*

$$t \in \{t_1 + (1/\beta)\ln(1/\delta), \ldots, t_2\}$$

*$\vec{p}^t$ is an $(\psi + \rho\delta)$-Nash equilibrium.*

**Proof:** As in Lemma 9, after $(1/\beta)\ln(1/\delta)$ steps we have $\|\vec{p}^t - \hat{a}(\mu_t)\|_\infty \leq \delta$. Therefore each player is playing a $O(\psi + \rho\alpha + \rho\delta)$-best response in $\hat{\mathcal{G}}$. □

Since Lemma 12 does not depend on the termination mechanism, Lemma 14 and the choice $\beta = \alpha/2$ implies the following theorem.

**Theorem 15** *For any $\delta > 0$, if* **SummLearn**(0) *is run for an infinite sequence of steps, the players play an $O(\psi + \rho\tau + \rho\delta + \rho\alpha)$-Nash equilibrium in all but at most $O((1/\alpha^2)\ln(1/\delta))$ steps.*

Note that though **SummLearn**(0) has no dependence on $\delta$ (only the global summarization mean must be broadcast), Theorem 15 provides a spectrum of statements about this algorithm parameterized by $\delta$ — as we reduce $\delta$, we give worse (larger) bounds on the total number of steps that a better approximation to equilibrium is played.